# Communication Interface Identifier Protocol (CIIP): An Energy Efficient Protocol for smaller IoT Sensor

Nasr Al-Zaben*, Md Mehedi Hassan Onik, Kim Chul-Soo, Yang Jinhong

ABSTRACT

Today we can use technologies like switched Ethernet, TCP/IP, high-speed wide area networks, and high-performance low-cost computers very easily. However, protocols designed for those communication are inefficient or not energy efficient. Smart home, smart grid, blockchain, Internet of Things (IoT) all these technologies are coming very rapidly with higher communication facilities demands an energy efficient Ethernet. Due to controller and network equipment use a huge quantity of energy. Layer to layer communication making our communication method more complex and costly. In this work, we propose an architecture, which will make the communication of sensor devices to outside world easier. Our proposed system removes certain layer from TCP-IP communication. We used a communication interface identifier protocol (CIIP) which can be used for smaller IoT sensors.

Keywords : IoT Sensor, Energy Efficient, Protocol, CIIP, Smart Grid Communication, Ethernet.

## Ⅰ. Introduction

Communication has great impact on power management, especially for real time communication smart grid is improving efficiency, security and reliability of current power sector. Internet of Things (ioT), 5G technology are blooming in a great rate. Security issues all over the world is increasing which also put pressure on smart grid also [1]. Smart grid facing implementation is a complex sector for its architecture, which ultimate demands a feasible and secure way of communication. Hackers are also attacking smart grid, for example, in 2015 the attack on Ukraine make the world think a lot for this sector where around 225000 people stayed without power, similarly in 2016 energy company call center was also under attack for stopping communication [2].

In this paper, we are proposing a protocol named Communication interface identifier protocol (CIIP). We introduce an identifier for a device, an adapter will attached with this device so that it can communicate with internet protocol (IP) and device identifier number. Our proposed protocol is mainly designed for slightly smaller IoT sensor devices.

One of the main problem to the deployment of Smart Grid applications is the limited capability of today's utility communication infrastructure in terms of scalability, reliability and security. The recent 2005 Houston blackout [3], illustrate the importance of a secure data sharing network. NASPInet is a framework for providing a robust and secure synchronized data sharing infrastructure for the interconnected North American electric power system [4]. Number of connected M2M devices and consumer electronics will surpass the number of human subscribers using mobile phones, personal computers, laptops and tablets by 2020 [5]. Moving forward, by 2024, the overall IoT industry is expected to generate a revenue of 4.3 trillion dollars [6] across different sectors such as device manufacturing, connectivity, and other value added services. Modern substation automation systems (SAS) are implementing switched Ethernet-based international communication standard IEC 61850 for achieving the smart grid goals [7]. IEEE 802.21 is also working on this the standardization of smart grid communication [8]. We can combine the existing wireless communication protocols into the following six standards: 1. Satellite 2. WiFi 3. Radio Frequency (RF) 4. RFID 5. Bluetooth / BLE 6. Near Field Communication (NFC). Above mentioned communication media uses completely depend on user's requirement and transmitted items. Since smart grid and IoT is producing lots of data every day, our protocols are also changing to keep pace with this data change. An overview is given bellow:

a) Infrastructure (ex: 6LowPAN, IPv4/IPv6, RPL)
b) Identification (ex: EPC, uCode, IPv6, URIs)
c) Communication / Transport (ex: Wifi, Bluetooth, LPWAN)
d) Discovery (ex: Physical Web, mDNS, DNS-SD)
e) Data Protocols (ex: MQTT, CoAP, AMQP, Websocket, Node)
f) Device Management (ex: TR-069, OMA-DM)
g) Semantic (ex: JSON-LD, Web Thing Model)
h) Multi-layer Frameworks (ex: Alljoyn, IoTivity, Weave, Homekit).

---

*Department of Computer Engineering, Inje University, Gimhae-50834, South Korea

When connecting a few hundred or even several thousand devices in about 10 models to the existing IoT Eco framework sufficient. But if this had grown to billions, tens of billions, more size is an integrated central system will cause a bottleneck. Feasibility of using IEEE 802.11ah for IoT/M2M use cases is studied in [9]. Low power consumption and low cost is an important aspect. Such MTC application are categorized as Massive MTC [10] in 5G network. IEEE is extending range and reducing power consumption of their 802.15.4 [11] and 802.21 [12] standards with the set of new specifications for the physical and the MAC layers. Energy saving of network equipment is a serious issue, which needs our full attention for lesser carbon emission were discussed in studies like [13-14].

## II. Proposed Methods

This section explains architecture for our proposed protocol, which is mainly feasible for comparatively smaller sensor devices. Lots of data loss occur unstable wireless environment; because the TCP protocol to send and receive a number of control packets to ensure reliable communication with network devices. Due to performance issues in WiFi environment, though the protocol itself is not responsible for retransmission but for upper layer issues causes lots of retransmission and which is inefficient. Therefore, In order to develop the low-power IoT device, we need to transform the Physical Layer ~ Network Layer protocol.

### 1. Communication interface identifier protocol (CIIP)

The Communication interface identifier protocol (CIIP's) Communication ID will be mapped into "IP: PORT" format through CIIP to IP adapter device. So, the end-point of CIIP does not need to implement upper layers such as IP, TCP. Maybe it will implement just CIIP as physical layer and data-link layer. And application layer for users. It does not have any complexity because the protocol specification does not need to perform much encapsulation before transmitting the packet with CIIP.

CIIP Networks are consisted with CIIP Adapter and CIIP End-point. Application cannot access other fields except payload. So, if application want to send the packet, it will just set payload through Memory Mapped IO and a few number of PIO (Port IO) ports. Required PIO port must be including "RETR (Retransmission required)" and "ACT (Act the sending operation)" and "PPD (Packet Pending for receiving)". Our main idea is less communication for less power consumption and IoT device that require low power needs more simplified and to simplify existing communication protocols in order to connect the device to network, so we are considering this specialized protocol as an alternative. In addition, they can communicate with each other, the packet with the Communication Interface Identifier (CID, communication interface Identifier). The router does not guarantee a strong computing power, like a PC or server environment, IoT It can handle the higher protocol more effective than implementing the IP protocol devices and the CIIP communication can be a good option that can be routed to the network environment.

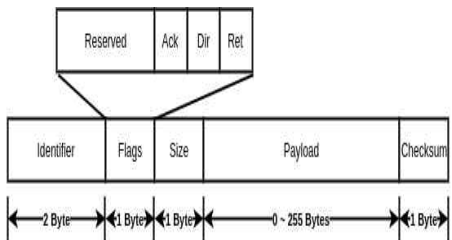

Fig. 1. Communication Interface Identifier Protocol (CIIP) packet format

Table 1. Frame format of Communication Interface Identification Protocol

| Name | Expression |
|---|---|
| Identifier | Dir == 0: the identifier for recognizing the sender. Dir == 1: the identifier for recognizing the receiver. |
| Dir Flag | Not set (0), this packet is from CIIP End-point. If set (1), this packet is from CIIP Adapter. |
| Ret Flag (Retransmission flag) | If set, previous was arrived well. |
| Ack Flag (Acknowledgment flag) | If set, previous was arrived well. This will be filled out with payload's length. |
| Size Checksum | The simple SUM of entire packet bytes. |

Above table (Table 1) give us idea how we can understand about the packet format and its meaning. Since our protocol is consisted with CIIP adapter and CIIP endpoint, from Identifier we can identify a packet, if it is from CIIP end point or CIIP adapter easily. Figure (Fig. 2) shows the current status of the protocol and we tried to remove those layer. In our proposed idea we inserted Hardware compatibility layer (HAL) is attached with CIIP protocol and every kind of information exchange is done between CIIP or cellular network and HAL after that it become more compatible with user's application layer. Which remove layer to layer barrier and energy consumption too.

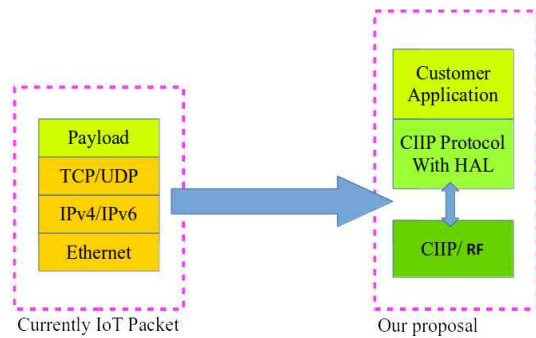

Fig. 2. OSI layer comparison

Figure below (Fig. 3) describe how IoT devices data with the help of middle edge RF-wireless router (light weight communication) reach external network. In our proposed design Customer application (user equipment) will receive the payload from external network using few obstacles.

Figure (Fig. 4) below describe how a packet to IP / Mobile-Network protocol work in L4 level in CIIP protocol, how it will route among themselves is shown in figure (Fig. 4). Each IoT smart sensor is a node which determine a destination to be reach. Packets are sent from each device by using the (Communication Identification number) CID of the device. Any particular the IoT device report on device location to its respected registration server and enable access from the outside network. In a regular interval each node or IoT sensor send HBP to the server connected with the network for verifying its location information, if sensor device stop sending this HBP information to the server, our proposed

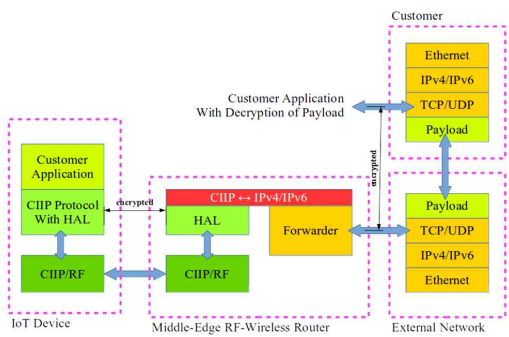

Fig. 3. Operational Architecture of CIIP

protocol will automatically recognize that sensor as a dead or unreachable sensing device, which result an automatic removal of that device from the server. Finally, the overhead of conventional LTE or other mobile protocol needed to connect to sensor for IoT technologies can minimize the power consumption of the device itself effectively.

Figure (Fig. 4) represent our adopter or repeater position and working architecture. Left side there are external network (i.e. 4G/LTE/5G),

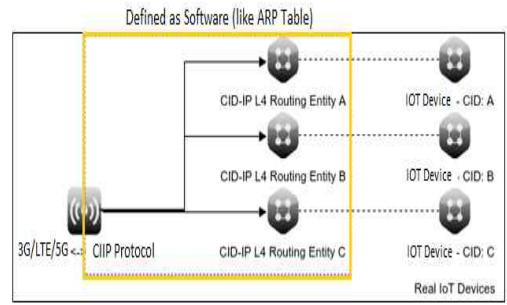

Fig. 4. CIIP adapter overview

which is directly connected with our proposed adopter for directing the packets towards and from IoT sensor devices. We put layer 4 switches which repeat packets according to the IoT device identification number (CID). Each IoT devices will have individual identification number, which will help L4 switch to identify its packet destination. This newly proposed method removed some layer from Open Systems Interconnection model (OSI model) and introduce device interface identifier, which can save energy as well as make smart grid and IoT communication method compatible. This can also fit with upcoming 5G technology, with some change in packet format it can be compatible with 5G technologies. A possible implementation architecture is given below in figure (Fig. 5):

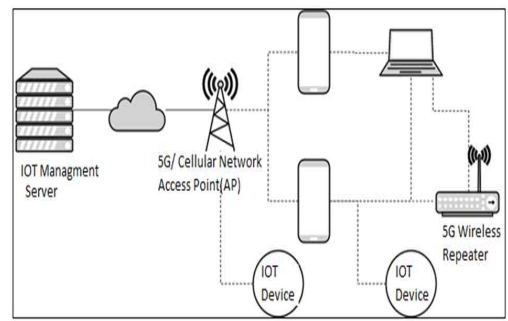

Fig. 5. CIIP and IoT sensor/ device overview

## III. Discussion

Current protocols available for collecting sensor information is complex and costly. We tried to propose a cost effective communication protocol of future technologies like IoT and smart grid. Though our proposed device is mostly feasible in light sensors but it can affect for collecting information from sensitive places. Already used protocol like IEC 61850 and IEEE 802.21 must improve with time since, with IoT and 5G technologies, world will have more device, which will ultimately put pressure on energy consumption and carbon production [7] [8]. Thus, we took the opportunity to present this protocol CIIP. Our future plan is to implement this with real life scenario.

## IV. Conclusions

In this paper, we tried to propose an energy efficient feasible communication protocol for Internet of things (IoT) sensors, which size is comparative lower. We all know that, energy consumption is big issue for smaller sensor devices, too many layers make this energy consumption even higher. If our proposed protocol is used with some adopter, can communicate with IP based devices too. In future, our goal is to implement this idea and bring a real representation and compare with currently used protocols. This approach can bring an easy communication protocol among IoT devices and smart grid too.